\documentclass[english]{article}
\usepackage{lmodern}
\usepackage[T1]{fontenc}
\usepackage[latin9]{inputenc}
\usepackage{amsmath}
\usepackage{amssymb}
\usepackage{graphicx}
\usepackage{babel}
\usepackage{geometry}
\usepackage{authblk}

\begin{document}

\title{Thermodynamics in the universe described by the emergence of the
space and the energy balance relation }
\author[a]{Fei-Quan Tu \thanks{Corresponding author: fqtuzju@foxmail.com}}
\author[b]{Yi-Xin Chen}
\author[a]{Qi-Hong Huang}
\affil[a]{School of Physics and Electronic Science, Zunyi Normal University, Zunyi 563006, China}
\affil[b]{Zhejiang Institute of Modern Physics, Zhejiang University, Hangzhou 310027, China}


\renewcommand*{\Affilfont}{\small\it} 
\renewcommand\Authands{ and } 
\date{} 

\maketitle
\begin{abstract}
It has previously been shown that it is more general to describe the
evolution of the universe based on the emergence of the
space and the energy balance relation. Here we investigate
the thermodynamic properties of the universe described by such a model.
We show that the first law of thermodynamics
and the generalized second law of thermodynamics
(GSLT) are both satisfied and the weak energy condition are also fulfilled
for two typical examples. Finally we examine the physical consistency
for the present model.
\end{abstract}

\section*{1. Introduction}

Numerous astro-observations show that our universe is in accelerating
expansion at present \cite{key-1,key-2}. Usually there are two ways
to explain the phenomenon if we consider the evolution of the universe
from the view of point of the dynamics of gravity. One is the modification
of geometric part of Einstein's field equation, such as $f(R)$ theory,
Lanczos-Lovelock gravity theory. The other is the modification of
material part of Einstein's field equation by introducing the extra
matter with negative pressure or the scalar field (called dark energy).
A good model which can explain the accelerating expansion of the universe
is $\mathrm{\Lambda CDM}$ (lambda cold dark energy) model which considers
a cosmological constant, i.e. the value of the equation of state parameter
$\omega=-1$ \cite{key-3,key-4}. However, astronomical observations
allow also $\omega$ to vary with time. The variation with time is
described usually by the slowly varying scalar field such as the quintessence
\cite{key-5,key-6,key-7,key-8,key-9}, kinetic energy driven k-essence\cite{key-10,key-11,key-12}
and tachyon \cite{key-13,key-14,key-15,key-16,key-17}. These models
describe the accelerating expansion of the universe well.

Studying gravity from a thermodynamic point of view is an interesting
field in modern theoretical physics. The deep connection between gravity
and thermodynamics is accepted generally because of the black hole
thermodynamics\cite{key-18,key-19,key-20} and Ads/CFT correspondence\cite{key-21}.
In 1995, Jocbson derived the Einstein equation from the proportionality
of entropy and horizon area together with the relation $\delta Q=TdS$
connecting heat flow $\delta Q$, entropy $S$ and Unruh temperature
$T$ \cite{key-22}. Frolov and Kofman \cite{key-23} found that one
of the Friedmann equations with the slow-roll scalar field can be
reproduced when they used the first law of thermodynamics $-dE=TdS$
where $dE$ is the amount of the energy flow through the quasi-de
Sitter apparent horizon, although the topology of quasi-de Sitter
apparent horizon is quite different from that of the local Rindler
horizon in Ref.\cite{key-22}. Danielsson \cite{key-24} derived the
Friedmann equations by calculating the heat flow through the horizon
in an expanding universe and by applying the relation $\delta Q=TdS$
to a cosmological horizon. Padmanabhan \cite{key-25}, in the reverse
way, showed that gravitational field equations in a wide variety of
theories can reduce to the thermodynamic identity $TdS=dE+PdV$ when
evaluated on a horizon. These conclusions further reveal the relationship
between the thermodynamics of horizon and the dynamics of gravity.
Furthermore, Padmanabhan \cite{key-26} revealed the relationship
between the degrees of freedom of space and the dynamic evolution
of the universe. He argued that the difference between the number
of the surface degrees of freedom $N_{sur}$ and that of the bulk
degrees of freedom $N_{bulk}$ in a region of space drives the accelerating
expansion of the universe through a simple equation $\Delta V=\Delta t\left(N_{sur}-N_{bulk}\right)$,
where $V$ is the Hubble volume in Planck units and $t$ is the cosmic
time in Planck units, and derived the standard Friedmann equation
of the FRW universe.

From the point of view of thermodynamics, it tends to spontaneously
thermodynamic equilibrium for an isolated thermodynamical system.
That is to say, the entropy of an isolated thermodynamical system,
$S$, cannot decrease and reaches its maximum finally. In the black
hole physics, there exists the similar law which is called as the
generalized second law of thermodynamics \cite{key-19}. The law states
that the sum of the entropy of the black hole horizon, $S_{h}$, plus
the entropy of the matter, $S_{m}$, cannot decrease with time, i.e.
$\dot{S}=\dot{S}_{m}+\dot{S}_{h}\geq0$ where the dot denotes the
derivative with respect to time. After that, the GSLT was extended
to the cosmological horizons \cite{key-27,key-28}. In the cosmological
context, the GSLT has been extensively studied (see, for example,
\cite{key-29,key-30,key-31}).

In Ref.\cite{key-32}, we considered the accelerating expansion
of the universe in the late time based on the emergence of space and
the energy balance relation $\rho V_{H}=TS$, where $\rho$ is the
energy density of the cosmic matter, $S=A_{H}/(4L_{p}^{2})=\pi H^{-2}/L_{p}^{2}$
is the entropy associated with the area of the Hubble sphere $V_{H}=\frac{4\pi}{3H^{3}}$
and $TS$ is the heat energy of the boundary surface. Then we
found that the obtained solutions of the evolution
of the universe include the solutions obtained from the standard general
relativity theory, and concluded that it is more general to describe
the evolution of the universe in the thermodynamic way. So it is interesting
to investigate whether the first law of thermodynamics and the GSLT
hold in the model described by Ref.\cite{key-32}.

The goal of the present paper is to study the thermodynamical behavior of the universe
considered in Ref.\cite{key-32} by means of its description of
the emergence of space and the energy balance relation.
Our analysis shows that the first law of thermodynamics is
satisfied and is actually Clausius relation. The validity of Clausius
relation means that the evolution of the universe can be deemed as
a series of quasistatic processes. We also show that the GSLT holds
in the total accelerating evolutionary history of the universe and the total entropy
of the universe tends to the maximal value when the universe
evolves to the de Sitter universe by considering two typical
examples.

The paper is organized as follows. In Sec. 2, we review
briefly the model which describes the evolution of the universe based
on the emergence of space and the energy balance relation. In Sec.
3, we show that the first law of thermodynamics holds in the universe
descried by the present model. In Sec. 4, the validity of the GSLT
and thermodynamic equilibrium are shown, we obtain also the constraints
imposed on the energy density and the pressure of the matter. Our
conclusions are presented in Sec. 5. We use units $\ensuremath{c=\hbar=1}$.

\section*{2. Dynamical evolution equations of the universe based on the emergence
of space and the energy balance relation}

Let us begin with the FRW metric which describes the homogeneous and
isotropic universe
\begin{equation}
ds^{2}=-dt^{2}+a^{2}(t)\left(\frac{dr^{2}}{1-kr^{2}}+r^{2}d\Omega^{2}\right)=h_{ab}dx^{a}dx^{b}+R^{2}d\Omega^{2},
\end{equation}
where $R=a(t)r$ is the comoving radius, $h_{ab}=diag\left(-1,\;\frac{a^{2}}{1-kr^{2}}\right)$
is the metric of 2-spacetime ($x^{0}=t,\; x^{1}=r$) and $k=0,\;\pm1$
denotes the curvature parameter. Padmanabhan \cite{key-26} thought that
our universe is asymptotically de Sitter and its evolution can be
described by the following law
\begin{equation}
\frac{dV}{dt}=L_{p}^{2}\left(N_{sur}-N_{bulk}\right),
\end{equation}
where $L_{p}$ is the Planck length,
\begin{equation}
N_{sur}=\frac{4\pi}{H^{2}L_{p}^{2}}
\end{equation}
is the number of the surface degrees of freedom on the Hubble horizon
in which $H$ is the Hubble constant,
\begin{equation}
N_{bulk}=\frac{|E|}{(1/2)k_{B}T}=-\frac{2(\rho+3p)V}{k_{B}T}
\end{equation}
is the number of the bulk degrees of freedom in which $T$ is the
temperature of the horizon, $k_{B}$ is the Boltzmann constant,
$|E|=|(\rho+3p)V|$ is the Komar energy
and $V=\frac{4\pi}{3H^{3}}$ is the Hubble volume. The law (2) indicates
that the difference between $N_{sur}$ and $N_{bulk}$ drives the
universe towards ``holographic equipartition'' (i.e. $N_{sur}=N_{bulk}$).

According to the analysis of Ref.\cite{key-32}, the temperature of
Hubble horizon of the flat FRW universe is employed as
\begin{equation}
T=\frac{H}{2\pi}\left(1+\frac{\dot{H}}{2H^{2}}\right).
\end{equation}
Here, we assume $1+\frac{\dot{H}}{2H^{2}}>0$. The study of quantum
field theory in a de Sitter space \cite{key-33} showed that a freely
falling observer would measure a temperature $T=H/2\pi$ on the de
Sitter horizon when the radius of the de Sitter horizon is taken as
$1/H$ . Our universe is not exactly de Sitter but asymptotically
de Sitter, so the temperature of the Hubble horizon should tend to
$H/2\pi$ when $t$ becomes large enough. In fact, the approximation
$\mid\dot{H}/2H^{2}\mid\ll1$ have been used in calculating the energy
flow crossing the apparent horizon \cite{key-23,key-24,key-34,key-35}.
Therefore, it seems to be reasonable to assume $1+\frac{\dot{H}}{2H^{2}}>0$
when we investigate the thermodynamic properties and dynamical behavior
of the universe. In Sec. 4, we will show the validity of the assumption
$1+\frac{\dot{H}}{2H^{2}}>0$ by two typical examples.

Thus, inserting Eq.(3), Eq.(4) and Eq.(5) in Eq.(2), we obtain the
Friedmann acceleration equation
\begin{equation}
\frac{\ddot{a}}{a}=H^{2}+\dot{H}=-\frac{4\pi L_{p}^{2}}{3}(\rho+3p)-\frac{\dot{H}}{2}-\frac{\dot{H^{2}}}{2H^{2}}.
\end{equation}
On the other hand, according to the energy balance relation $\rho V_{H}=TS$
\cite{key-36}, we obtain another evolution equation of the universe
\begin{equation}
H^{2}=\frac{8\pi L_{p}^{2}}{3}\rho-\dot{\frac{H}{2}}.
\end{equation}
Combining Eq.(6) and Eq.(7), the following equation
\begin{equation}
\dot{\rho}+3H(\rho+p)=\frac{3}{8\pi L_{p}^{2}}\left(\frac{\ddot{H}}{2}-\frac{\dot{H}^{2}}{H}\right)
\end{equation}
can be obtained. In this way, we obtain the dynamical evolution equations
of the universe based on the emergence of space and the energy balance
relation. It was shown \cite{key-32} that it is more general to describe
the evolution of the universe in such a thermodynamic way because
the solutions of the dynamical evolution equations in such a model
include the solutions obtained from the standard general relativity
theory.

\section*{3. First law of thermodynamics for the present model}

Now that it is more general to describe the evolution of the universe
from a thermodynamic perspective, then we would like to ask whether
the first law of thermodynamics and the GSLT can hold in such a model.
Furthermore, we may ask what are the constraints on the evolution
of the universe if the GSLT holds. In this and next sections, we will
discuss the first law of thermodynamics and the GSLT in such a model
respectively.

The study of quantum field theory in a de Sitter space \cite{key-33}
showed that a freely falling observer would measure a temperature
$T=\kappa/2\pi$ on the de Sitter horizon where $\kappa$ is the surface
gravity. For the Q space, Bousso \cite{key-37} argued its thermodynamical
description and showed that the first law of thermodynamics $-dE=TdS$
holds on the apparent horizon. Furthermore, Cai and Kim \cite{key-34}
derived the Friedmann equation of the FRW universe with any spatial
curvature based on the first law of thermodynamics. Whether the first
law of thermodynamics holds on the horizons in different gravity theories
have been studied generally (For example, see, \cite{key-38,key-39,key-40}).

Now let us show that the first law of thermodynamics holds in the
universe described by the present model. The amount of energy crossing
the Hubble horizon during the time interval $dt$ \cite{key-37,key-41}
is
\begin{equation}
-dE=4\pi R_{h}^{2}T_{\mu\nu}k^{\mu}k^{\nu}dt=\frac{4\pi}{H^{2}}(\rho+p)dt,
\end{equation}
where $R_{h}$ is the Hubble radius and $k^{\mu}$ is the future directed
ingoing null vector field.

Using Eq.(6) and Eq.(7), we obtain
\begin{equation}
\rho+p=-\frac{\dot{H}}{4\pi L_{p}^{2}}\left(1+\frac{\dot{H}}{2H^{2}}\right),
\end{equation}
so the amount of energy crossing the Hubble horizon during the infinitesimal
time interval is expressed as
\begin{equation}
-dE=-\frac{\dot{H}}{H^{2}L_{p}^{2}}\left(1+\frac{\dot{H}}{2H^{2}}\right)dt.
\end{equation}

On the other hand, we can obtain
\begin{equation}
TdS=-\frac{\dot{H}}{H^{2}L_{p}^{2}}\left(1+\frac{\dot{H}}{2H^{2}}\right)dt,
\end{equation}
where we use the definition of temperature Eq.(5) and the area-entropy
relation $S=$$\frac{A}{4L_{p}^{2}}=\frac{\pi}{H^{2}L_{p}^{2}}$.
Comparing Eq.(11) with Eq.(12), we obtain the following equality
\begin{equation}
-dE=TdS,
\end{equation}
which implies that the first law of thermodynamics holds in the present
model.

It is important to note that the strong energy condition $\rho+3p\geq0$
is broken from Eq.(4). However, we can see that the null energy condition
$\rho+p\geq0$ can be satisfied when $\dot{H}$ is nonpositive from
Eq.(10) because the term $1+\frac{\dot{H}}{2H^{2}}$ which is related
with the temperature of the Hubble horizon is positive. In fact, it
satisfies the GSLT for the accelerating universe which satisfies the
null energy condition.

It is also worth mentioning that the amount of heat flux crossing
the Hubble horizon during the infinitesimal time interval, $\delta Q$
, is the change of the energy inside the Hubble horizon $-dE$. The
minus appears due to the fact that the energy inside the Hubble horizon
decreases when the heat flux flows out of the Hubble horizon, so the
law (13) is actually Clausius relation $\delta Q=TdS$. Therefore,
the evolution of the universe in the present model can be deemed as
a series of quasistatic processes because Clausius relation works
only when the thermodynamic process is reversible. Thus, the temperature
of the matter inside the Hubble horizon can be taken as the temperature
of the Hubble horizon. This is an important relation that we will
use when we discuss the GSTL in the next section.

\section*{4. Validity of the GSLT and thermodynamic equilibrium}

We have shown that the first law of thermodynamics holds on the horizon
in the previous section, it is natural to ask if the GSLT holds in
such a model. In the cosmological context, the GSLT denotes that the
sum of the entropy of the cosmological horizon, $S_{h}$, plus the
entropy of the matter inside the horizon, $S_{m}$, is a nondecreasing
function. That is to say, the GSLT can be formulated as \cite{key-27,key-28,key-42}
\begin{equation}
\dot{S}=\dot{S}_{m}+\dot{S}_{h}\geq0.
\end{equation}
Further, if the universe can reach an equilibrium state eventually,
then the total entropy must satisfy the inequality
\begin{equation}
\ddot{S}=\ddot{S}_{m}+\ddot{S}_{h}\leq0
\end{equation}
at least at the last stage of evolution. The physical meaning of this
inequality can be explained by the fact that the entropy of the universe
increases less and less and reaches a maximum if the universe reaches
an equilibrium state eventually.

Here we would like to point out that the formula (15) is slightly
different from the one in the references \cite{key-43,key-44} where
the authors used the expression that the second derivative of the
total entropy is less than $0$, i.e. $\ddot{S}=\ddot{S}_{m}+\ddot{S}_{h}<0$.
However, we allow the equal sign of the inequality (15) to be valid
because it is possible for the total entropy to take the maximum value
even if the first and second derivatives of the total entropy are
both zero. In fact, the first and second derivatives of the total
entropy are both zero when the universe evolves into the de Sitter
universe.

According to the Gibbs relation and the conclusion of the previous
section that the evolution of the universe can be deemed as a series
of quasistatic processes, we know that the matter of the universe
satisfies \cite{key-45,key-46,key-47}
\begin{equation}
TdS_{m}=d(\rho V)+pdV=(\rho+p)dV+Vd\rho,
\end{equation}
where $V=\frac{4\pi}{3H^{3}}$ is the Hubble volume and $T$ equals
to the temperature of the Hubble horizon. Substituting Eq.(5), Eq.(7)
and Eq.(10) into Eq.(16), we obtain the change rate of the entropy
of the matter
\begin{equation}
\dot{S}_{m}=\frac{2\pi}{H^{2}L_{p}^{2}\left(2H^{2}+\dot{H}\right)}\left[\frac{\ddot{H}}{2}+2H\dot{H}+\frac{\dot{H}^{3}}{H^{3}}+\frac{2\dot{H}^{2}}{H}\right].
\end{equation}
For the Hubble horizon, the change rate of the entropy is
\begin{equation}
\dot{S}_{h}=-\frac{2\pi\dot{H}}{H^{3}L_{p}^{2}}.
\end{equation}
Therefore, we can get the first derivative of the total entropy
\begin{equation}
\dot{S}=\dot{S}_{m}+\dot{S}_{h}=\frac{2\pi}{H^{3}L_{p}^{2}\left(2H^{2}+\dot{H}\right)}\left[\frac{\ddot{H}H}{2}+\frac{\dot{H}^{3}}{H^{2}}+\dot{H}^{2}\right]
\end{equation}
and the second derivative of the total entropy
\begin{equation}
\ddot{S}=\ddot{S}_{m}+\ddot{S}_{h}=\frac{\pi}{L_{p}^{2}}\left[\frac{-34H^{2}\dot{H}^{4}-10\dot{H}^{5}+12H^{3}\dot{H}^{2}\ddot{H}+4H\dot{H}^{3}\ddot{H}+2H^{6}\dddot{H}-H^{4}\left(20\dot{H}^{3}+\ddot{H}^{2}-\dot{H}\dddot{H}\right)}{H^{6}\left(2H^{2}+\dot{H}\right)}\right].
\end{equation}

Before we discuss the GSLT, let us obtain the constraints which are
imposed on the energy density and pressure of the matter by the present model.
The equation (10) can be transformed into
\begin{equation}
\dot{H}^{2}+2H^{2}\dot{H}+8\pi L_{p}^{2}H^{2}(\rho+p)=0.
\end{equation}
Solving this equation, we obtain the solutions
\begin{equation}
\dot{H}=-H^{2}\pm H\sqrt{H^{2}-8\pi L_{p}^{2}(\rho+p)},
\end{equation}
which imply that the sum of the energy density and the pressure must
satisfy the relation
\begin{equation}
\rho+p\leq\frac{H^{2}}{8\pi L_{p}^{2}}.
\end{equation}
This constraint gives the upper bound of the sum of the energy density
and the pressure.

Now we investigate the GSLT in the present time and the last time
of the evolution, respectively.

(i). The GSLT in the present time of the evolution. At the present
time, we assume that the scale factor behaves as
\begin{equation}
a(t)\propto t^{\alpha}
\end{equation}
where $\alpha$ is a constant greater than unity because the universe
is in accelerating expansion. This form can be obtained when the relation
$8\pi L_{p}^{2}\left(\rho+p\right)\propto H^{2}$ is satisfied. In Ref.\cite{key-32},
it has been proved that the form of the scale factor is $a(t)=t^{\frac{2}{3(1+\omega)}}$
if the equation of state of the matter is assumed as $p=\omega \rho$ where $\omega$ is
a constant that is not equal to $-1$.
In fact, a large number of papers on the accelerating expansion of the
universe have assumed that the scale factor is the form (24). For
example, the authors have pointed out that the rate of growth $a(t)\propto t^{2}$
is consistent with supernova observations in the Ref.\cite{key-48}.
After some calculations we obtain
\begin{equation}
H=\frac{\alpha}{t},\quad\dot{H}=-\frac{\alpha}{t^{2}},\quad\ddot{H}=\frac{2\alpha}{t^{3}}.
\end{equation}
Inserting Eq.(25) in Eq.(19), we obtain the change rate of the total
entropy
\begin{equation}
\dot{S}=\frac{2\pi t}{\alpha^{3}L_{p}^{2}}
\end{equation}
which is greater than zero obviously, so the GSLT is satisfied in
the present time of the evolution. The term related with the temperature
of the Hubble horizon, $1+\frac{\dot{H}}{2H^{2}}$, could be derived
from Eq.(25) as $1-\frac{1}{2\alpha}$, which shows that the temperature
of the Hubble horizon is positive. Further, we find that the null energy condition
holds in the present time of the evolution because Eq.(10) is positive.

(ii). The GSLT in the last time of the evolution. The Friedmann acceleration
equation (6) is derived from the fact that our universe is asymptotically
de Sitter, so the scale factor $a(t)\rightarrow Ae^{H_{0}t}$ when
time $t\rightarrow\infty$ where $A$ and $H_{0}$ are both positive
constants. Thus the scale factor can be taken as
\begin{equation}
a(t)\propto\sinh(H_{0}t).
\end{equation}
Under this assumption, we obtain the following physical quantities
\begin{equation}
H=\frac{\dot{a}}{a}=H_{0}\coth(H_{0}t)
\end{equation}
and
\begin{equation}
\dot{H}=-H_{0}^{2}\mathrm{csch^{2}}(H_{0}t),\quad\ddot{H}=2H^{3}\coth(H_{0}t)\mathrm{csch^{2}}(H_{0}t).
\end{equation}
Inserting Eq.(28) and Eq.(29) in Eq.(19), we obtain
\begin{equation}
\dot{S}=\frac{\pi(8\cosh(2H_{0}t)+\cosh(4H_{0}t)-1)\mathrm{\mathrm{sech}}^{4}(H_{0}t)\mathrm{sech}(2H_{0}t)\tanh(H_{0}t)}{4H_{0}L_{p}^{2}}\geq0,
\end{equation}
which implies that the total entropy is nondecreasing and the GSLT
is satisfied. The derivative of the above expression, i.e. the second
derivative of the total entropy is
\begin{equation}
\ddot{S}=-\frac{\pi(54\cosh(2H_{0}t)-52\cosh(4H_{0}t)+10\cosh(6H_{0}t)+\cosh(8H_{0}t)-45)\mathrm{\mathrm{sech}}^{6}(H_{0}t)\mathrm{sech}^{2}(2H_{0}t)}{16L_{p}^{2}}.
\end{equation}
Analyzing the expression (31), we obtain the conclusion $\ddot{S}\leq0$
for the sufficiently large time $t$ which implies that the universe
will tend to thermodynamic equilibrium. In order to see the conclusion
$\dot{S}\geq0$ and $\ddot{S}\leq0$ clearly, we draw Figure 1 and Figure 2
to show the variation of the first and second derivatives of the total entropy
in the time range of $1/H_{0}$ to $6/H_{0}$, respectively. The term
related with the temperature of the Hubble horizon, $1+\frac{\dot{H}}{2H^{2}}$,
could be derived from Eq.(28) and Eq.(29) as $\frac{1}{2}\left[1+\tanh^{2}(H_{0}t)\right]$.
This term is positive so the temperature of the Hubble horizon is
positive. Further, we obtain the conclusion from Eq.(10) that the
null energy condition holds in the last time of the evolution.

\begin{figure}[htbp]
\begin{minipage}[t]{0.45\linewidth}
\centering
\includegraphics[height=4.5cm,width=6cm]{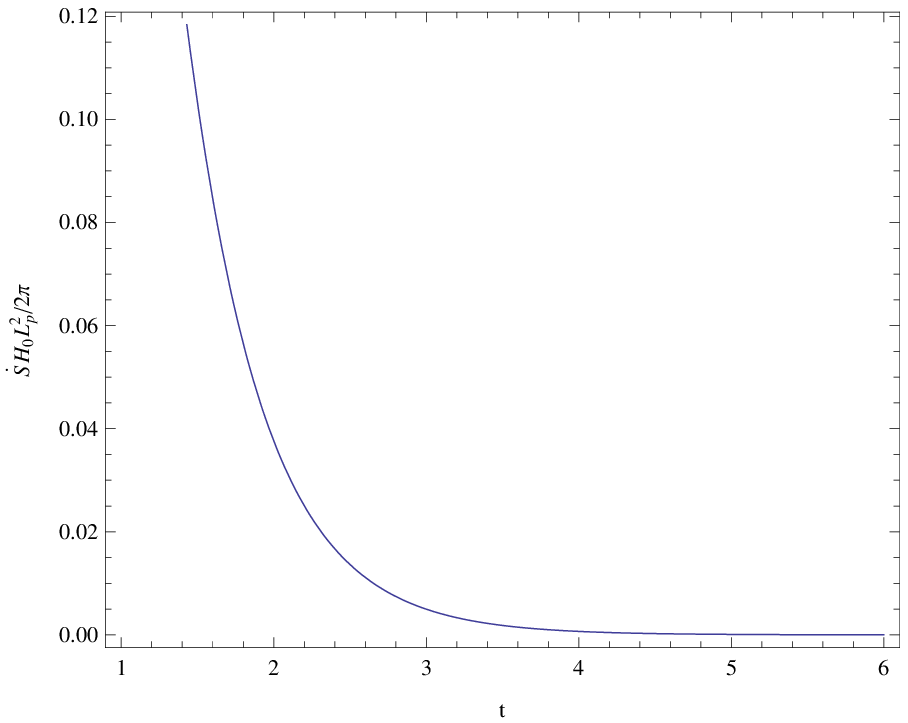}
\caption{The variation of the first derivative of the total entropy in the time
range of $1/H_{0}$ to $6/H_{0}$. This figure shows that the total
entropy is nondecreasing and the GSLT is satisfied.}
\end{minipage}%
\hfill
\begin{minipage}[t]{0.45\linewidth}
\centering
\includegraphics[height=4.5cm,width=6cm]{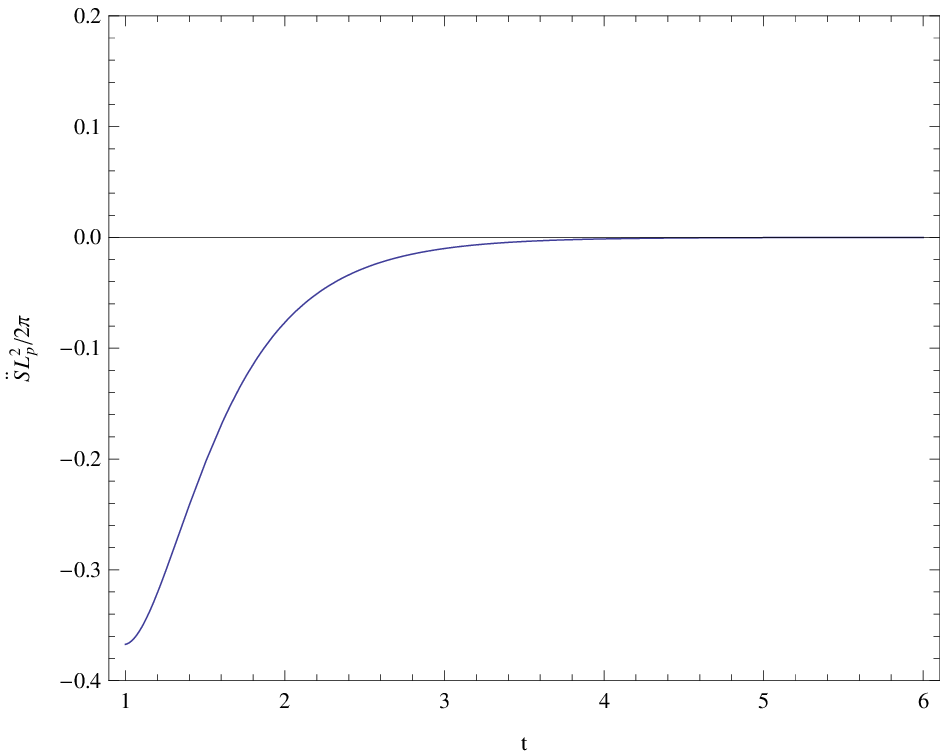}
\caption{The variation of the second derivative of the total entropy in the
time range of $1/H_{0}$ to $6/H_{0}$. This figure shows that the
universe will tend to thermodynamic equilibrium for the sufficiently
large time $t$.}
\end{minipage}
\end{figure}

At the end of this section, we investigate the special solution $\rho+p=0$
which depicts the de Sitter universe. When the equality $\rho+p=0$
is satisfied, we can obtain the solutions $\dot{H}=0$ or $\dot{H}=-2H^{2}$.
For the solution $\dot{H}=-2H^{2}$, we see $a(t)\propto t^{1/2}$ which implies
that the universe is not in accelerating expansion. This is inconsistent
with the current assumption. Hence the unique solution is $\dot{H}=0$ which implies
that $H$ is a constant for the de Sitter universe. Substituting the
solution $\dot{H}=0$ into Eq.(19) , we obtain $\dot{S}=0$ which
implies that the GSLT is satisfied for the de Sitter universe. According
to the above analysis, we conclude that the total accelerating evolutionary process
of the universe satisfies the GSLT and the total entropy of the universe
tends to the maximal value which equals to the total entropy
of the de Sitter universe in the present model.

\section*{5. Conclusions}

In this paper, we study the first law of thermodynamics and the GSLT
in the universe described by the emergence of space and the energy
balance relation. First, we obtain the evolution equations of the
universe based on the emergence of the space and the energy balance relation.
In the process of derivation of th\emph{}e temperature of the Hubble horizon, we assume that the term
$1+\dot{H}/H^{2}$ must be greater than zero. This assumption is reasonable
because this term equals exactly to unity for the de Sitter universe
and our universe is asymptotically de Sitter. Indeed, we show the
validity of the assumption for the accelerating universe whose evolution
law is $a(t)\propto t^{\alpha}$ or $a(t)\propto\sinh(H_{0}t)$ in
the section 4. Next, we show that the first law of thermodynamics
$-dE=TdS$ is satisfied for the present model. In fact, the validity
of the first law of thermodynamics implies the Clausius relation $\delta Q=TdS$
is satisfied in the present cosmological context. Therefore, the temperature
of the matter inside the universe can been taken as the temperature
of the Hubble horizon because the Clausius relation applies only to
variations between the nearby states of local thermodynamic equilibrium.

Then, we analyze the GSLT and get the change rate of the total entropy
according to the Gibbs relation and the area-entropy relation. Furthermore,
we obtain the constraints which are imposed on the energy
density and pressure of the matter by the present model.
These constraints are $\rho+p\leq\frac{H^{2}}{8\pi L_{p}^{2}}$
and $\rho+3p<0$ respectively. To arrive at more specific results,
we consider two typical examples in which the scale factor is taken
as $a(t)\propto t^{\alpha}$ and $a(t)\propto\sinh\left(H_{0}t\right)$.
The choice of the scale factor is based on the astronomical observation
and the consistency with the current model. Whether the scale factor
is taken as $a(t)\propto t^{\alpha}$ or $a(t)\propto\sinh\left(H_{0}t\right)$,
the GSLT and these constraints are satisfied. At the same time, the
null energy condition $\rho+p\geq0$ is also satisfied. In addition,
we find that the universe will reach a thermodynamic equilibrium
state and the total entropy reaches a maximal value when time $t$
tends to infinity.

Finally we must point out that these evolution equations have been
obtained and the dynamical properties of such an universe have been
studied in Ref.\cite{key-32}. However, here we analyze the thermodynamic
properties for this universe and find that the first law of thermodynamics
and the GSLT are satisfied for two typical examples. The conclusions
presented here further support the thermodynamic interpretation of
gravity and reveal the connection between gravity and thermodynamics.

\section*{6. Acknowledgments}
This work is supported by Doctoral Foundation of Zunyi Normal University(Grant No.BS[2016]03), Education Department Foundation of Guizhou Province(Grant No.QianjiaoheKYzi[2017]247) and the NNSF of China(Grant No.11775187).


\begin{thebibliography}{10}
\bibitem[1]{key-1} A. G. Riess et al., Astron. J. 116, 1009 (1998)

\bibitem[2]{key-2} S. Perlmutter et al., Astrophys. J. 517, 565 (1999);
598,102 (2003)

\bibitem[3]{key-3} T. Padmanabhan, Phys. Rept. 380, 235 (2003)

\bibitem[4]{key-4} P. J. E. Peebles, B. Ratra, Rev. Mod. Phys. 75,
559 (2003)

\bibitem[5]{key-5} B. Ratra, P. J. E. Peebles, Phys. Rev. D 37, 3406
(1988)

\bibitem[6]{key-6} P. G. Ferreira, M. Joyce, Phys. Rev. Lett. 79,
4740 (1997)

\bibitem[7]{key-7} E. J. Copeland, A. R. Liddle, D. Wands, Phys.
Rev. D 57, 4686 (1998).

\bibitem[8]{key-8} R. R. Caldwell, R. Dave, P. J. Steinhardt, Phys.
Rev. Lett. 80, 1582 (1998)

\bibitem[9]{key-9} I. Zlatev, L. M. Wang, P. J. Steinhardt, Phys.
Rev. Lett. 185, 896 (1999)

\bibitem[10]{key-10} T. Chiba, T. Okabe, M. Yamaguchi, Phys. Rev.
D 62, 023511 (2000)

\bibitem[11]{key-11} C. Armendariz-Picon, V. F. Mukhanov, P. J. Steinhardt,
Phys. Rev. Lett. 85, 4438 (2000)

\bibitem[12]{key-12} C. Armendariz-Picon, V. F. Mukhanov, P. J. Steinhardt,
Phys. Rev. D 63, 103510 (2001)

\bibitem[13]{key-13} G. W. Gibbons, Phy. Lett. B 537, 1 (2002)

\bibitem[14]{key-14} T. Padmanabhan, Phys. Rev. D 66, 021301(R) (2002)

\bibitem[15]{key-15} M. Fairbairn, M.H.G. Tytgat, Phy. Lett. B 546,
1 (2002)

\bibitem[16]{key-16} J. S. Bagla, H. K. Jassal and T. Padmanabhan,
Phy. Rev. D 67, 063504 (2003)

\bibitem[17]{key-17} E. J. Copeland, M. R. Garousi, M. Sami, S. Tsujikawa,
Phy. Rev. D 71, 043003 (2005)

\bibitem[18]{key-18} J. M. Bardeen, B. Carter and S. W. Hawking,
Comm, Math. Phys. 31, 161 (1973)

\bibitem[19]{key-19} J. D. Bekenstein, Phys. Rev. D 7, 2333 (1973)

\bibitem[20]{key-20} S. W. Hawking, Comm. Math. Phys. 43, 199 (1975)

\bibitem[21]{key-21} J. M. Maldacena, Adv. Theor. Math. Phys. 2,
231 (1998)

\bibitem[22]{key-22} T. Jacobson, Phys. Rev. Lett. 75, 1260 (1995)

\bibitem[23]{key-23} A. V. Frolov, L. Kofman, JCAP 0305, 009 (2003)

\bibitem[24]{key-24} U. H. Danielsson, Phys.Rev. D, 71, 023516 (2005)

\bibitem[25]{key-25} T. Padmanabhan, Class. Quan. Grav. 19, 5387
(2002); Phys. Reports, 406, 49 (2005); AIP Conference Proceedings,
989 114 (2007); Gen. Rel. Grav., 40, 2031 (2008)

\bibitem[26]{key-26} T. Padmanabhan, arXiv:1206.4916; Res. Astron.
Astrophys. 12, 891 (2012)

\bibitem[27]{key-27} P. C. W. Davies, Class. Quan. Grav. 4, L225
(1987)

\bibitem[28]{key-28} P. C. W. Davies, Class. Quan. Grav. 5, 1348
(1987)

\bibitem[29]{key-29} H. M. Sadjadi, Phys. Rev. D 73, 063525 (2006)

\bibitem[30]{key-30} A. Sheykhi, B. Wang, Phys. Lett. B 678, 434
(2009)

\bibitem[31]{key-31} A. Sheykhi, Eur. Phys. J. C 69, 265(2010)

\bibitem[32]{key-32} F. -Q Tu, Y. -X Chen, B. -Sun, Y. -C Yang, Phys. Lett. B 784, 411(2018)

\bibitem[33]{key-33} G. W. Gibbons, S. W. Hawking, Phys. Rev. D 15,
2738 (1977)

\bibitem[34]{key-34} R. -G. Cai, S. P. Kim, JHEP 02, 050 (2005)

\bibitem[35]{key-35} G. Calcagni, JHEP 09, 060 (2005)

\bibitem[36]{key-36} T. Padmanabhan, Comptes Rendus Physique 18,
275 (2017)

\bibitem[37]{key-37} R. Bousso, Phys. Rev. D 71, 064024 (2005)

\bibitem[38]{key-38} M. Akbar, R. -G. Cai, Phys. Lett. B 635, 7 (2006)

\bibitem[39]{key-39} M. Akbar, R. -G. Cai, Phys. Lett. B 648, 243
(2007)

\bibitem[40]{key-40} S. -F. Wu, B. Wang, G. -H. Yang, Nucl. Phys.
B 799, 330 (2008)

\bibitem[41]{key-41} S. Chakraborty, Phys. Lett. B 718, 276 (2012)

\bibitem[42]{key-42} R. Easther, D. Lowe, Phys. Rev. Lett. 82, 4967
(1999)

\bibitem[43]{key-43} D. Pavon, W. Zimdahl, Phys. Lett. B 708, 217
(2012)

\bibitem[44]{key-44} S. Saha, S. Chakraborty, Phys. Lett. B 717,
319 (2012)

\bibitem[45]{key-45} G. Izquierdo, D. Pavon, Phys. Lett. B 633, 420
(2006)

\bibitem[46]{key-46} A. Sheykhi, Eur. Phys. J. C 69, 265 (2010)

\bibitem[47]{key-47} A. Sheykhi, B. Wang, Phys. Lett. B 678, 434
(2009)

\bibitem[48]{key-48} T. Padmanabhan, T. R. Choudhury, Phys. Rev.
D 66, 081301(R) (2002)\end{thebibliography}
\end{document}